\documentclass[%
 aip,
 amsmath,amssymb,
 floatfix,
 reprint,%
]{revtex4-1}

\usepackage{graphicx}
\usepackage{dcolumn}
\usepackage{bm}

\usepackage[utf8]{inputenc}
\usepackage[T1]{fontenc}
\usepackage{mathptmx}
\usepackage{etoolbox}
\usepackage{physics}

\makeatletter
\def\@email#1#2{%
 \endgroup
 \patchcmd{\titleblock@produce}
  {\frontmatter@RRAPformat}
  {\frontmatter@RRAPformat{\produce@RRAP{*#1\href{mailto:#2}{#2}}}\frontmatter@RRAPformat}
  {}{}
}%
\makeatother
\begin{document}

\preprint{AIP/123-QED}

\title{Excited-State Downfolding Using Ground-State Formalisms}
\author{Nicholas P. Bauman}
\email{nicholas.bauman@pnnl.gov}
 \affiliation{Physical Sciences Division, Pacific Northwest National Laboratory, Richland, WA, 99352, USA}

\date{\today}

\begin{abstract}
Downfolding coupled cluster (CC) techniques are powerful tools for 
reducing the dimensionality of many-body quantum problems. This work 
investigates how ground-state downfolding formalisms can target excited 
states using non-Aufbau reference determinants, paving the way for applications of quantum computing in excited-state chemistry. This study focuses on 
doubly excited states for which canonical equation-of-motion CC 
approaches struggle to describe unless one includes higher-than-double 
excitations. The downfolding technique results in state-specific effective 
Hamiltonians that, when diagonalized in their respective active spaces, 
provide ground- and excited-state total energies (and therefore 
excitation energies) comparable to high-level CC methods. The performance 
of this procedure is examined with doubly 
excited states of H$_{2}$, Methylene, Formaldehyde, and Nitroxyl.

\end{abstract}

\maketitle

\section{\label{sec:Intro} Introduction}

Calculating excited states remains an important challenge in quantum chemistry despite over a decade of strong theoretical development. The many reasons why it is challenging to describe excited states are compounding. First, there is the chemical nature of the problem, which includes the range of excitation energies, the density of states, and the nature of the excited state, to name a few factors.\cite{Vogler2001, SerranoAndres} Then there is the task of determining an appropriate methodology to accurately describe the state of interest. For higher-than-singly excited states, reliable high-level theories are required to accurately capture these states, which can be prohibitively expensive. After determining a desirable methodology, one is subjected to the available eigensolver algorithms and their corresponding operation count, memory requirements, and convergence issues. Given these hurdles, it seems fitting that excited state calculations can greatly benefit from incorporating modern technology and unconventional approaches. 

In chemistry, quantum computing has the potential to revolutionize the field by enabling the simulation of molecular systems that are beyond the capabilities of classical computers.\cite{aspuru2005simulated, Aspuru2019, Aspuru2020, Claudino-2022} The development of quantum algorithms has been primarily and largely devoted to describing the ground state. However, there are a handful of algorithms for excited states. The quantum phase estimation (QPE) algorithm\cite{nielsen2010quantum,kitaev1995quantum} is a powerful tool for calculating excited states, as the probability of sampling a particular state is directly proportional to the square of the overlap of the target state and a trial state. In earlier work, we demonstrated the ease of use of the QPE algorithm in calculating a variety of high-energy excited states using simple postulated initial states.\cite{bauman2021corelevel} The variational quantum eigensolver, usually applied to ground-state problems, can be extended to calculate excited states through the multistate-contracted,\cite{parrish_mcvqe,parrish2019hybrid} folded spectrum,\cite{mcclean_folded,thom_folded} and Lagrangian-based approaches.\cite{kuroiwa2021penalty,greene2021generalized,selvarajan2022variational} Other methods for excited states involve the imaginary-time variational quantum simulator,\cite{Aspuru-RSC,mcardle2019variational,motta2020determining} Peeters--Devreese--Soldatov energy functional,\cite{peeters1984upper,soldatov1995generalized} quantum subspace expansion,\cite{mcclean2017hybrid,urbanek2020chemistry} and the quantum equation-of-motion approach,\cite{ollitrault2020quantum} to name a few. However, until quantum algorithms and hardware mature, the application of these techniques is greatly limited. In order to perform meaningful quantum calculations, methods for calculating excited states must be coupled with formalisms to reduce the dimensionality of the problem (see discussion in Ref. \onlinecite{DUCC_dynamics}).

Mathematically rigorous formulations for reducing the dimensionality/cost of quantum formulations are necessary for expanding the envelope of system sizes tractable with current and near-term quantum simulators and hardware. There have been a variety of approximations and techniques in recent years for reducing the dimensionality and the complexity of quantum calculations.\cite{transcorrelated1,transcorrelated2,transcorrelated3,covos1,liu2022reducing,kowalski2022fock,callahan2021dynamical,fan2023quantum,huang2023leveraging} One of the most promising formalisms is the downfolding technique based on the double unitary coupled cluster (DUCC) ansatz,\cite{DUCC1, DUCC_VQE, DUCC_dynamics, Bauman_ExcitedDUCC, chladek2021variational, DUCC_DOUBLE,bauman2022coupled} which constructs effective (or downfolded) Hamiltonians in a small-dimensionality sub-space of the entire Hilbert space, which is commonly defined as an active space. The resulting downfolded Hamiltonians integrate out the external (out-of-active-space) Fermionic degrees of freedom from the internal (in-the-active-space) parameters of the wave function, which can be determined as components of the eigenvectors of the downfolded Hamiltonians in the active space. 

In earlier work, we introduced an extension of the DUCC approach for excited states.\cite{Bauman_ExcitedDUCC} The method combined excited-state equation-of-motion CC (EOM-CC) theory operators with ground-state cluster operators to construct a state-selective downfolded Hamiltonian for the targeted excited state, which worked well for singly excited states. Since this approach requires an underlying EOM-CC calculation, it is computationally expensive to extend to describe higher-than-singly excited states and subject to all the hurdles described earlier. So, it would seem beneficial to circumvent any EOM-CC calculation and find a technique that utilizes the well-defined ground-state downfolding procedure to construct excited-state effective Hamiltonians.

One of the simplest yet attractive approaches to approximating excited states has been to use self-consistent field (SCF) solutions.\cite{deltascf5,deltascf1,deltascf2,deltascf3,deltascf4,deltascf6} In this way, excitation energies are given by the energy difference between two distinct SCF solutions. This approach, often referred to as the $\Delta$-SCF approximation, brings the advantage of orbital relaxation. One can also improve the electron-electron correlation effects by carrying out many-body approaches on the distinct SCF solutions to build on these solutions. This means robust high-level ground-state methodologies can be applied to study excited states while avoiding all of the stumbling blocks of conventional excited-state many-body approaches.

The single-reference coupled cluster approach\cite{cc1,cc2,cc3,cc4,cc5,cc6} is well known for accurately describing ground state properties in molecular systems. However, the nature of the non-linear CC equations is sometimes undervalued and underutilized. One of the appealing features of the CC equations is the ability to converge to excited states with the same symmetry as the reference function.\cite{monkhorstcicc} Convergence to a targeted solution is aided by starting with an SCF reference solution corresponding to a targeted state. The difference between ground- and excited-state solutions define the $\Delta$-CC approach, which has been strongly demonstrated and investigated.\cite{ccsolutions1,ccsolutions2,ccsolutions3,ccsolutions4,ccsolutions5,ccsolutions6,ccsolutions7,ccsolutions8,ccsolutions9,ccsolutions10,ccsolutions11,ccsolutions12,ccsolutions13,lee2019excited} It is also worth mentioning that the nature of the CC equations to converge to excited states has been studied in the context of multi-reference CC methods which has lead to the understanding of the 'intruder' state problem.\cite{paldus1989spin}

Starting with an SCF solution for an excited state, one can utilize ground-state downfolding techniques based on the DUCC formalism to construct a state-specific effective Hamiltonian for the target excited state. In this work, I investigate the utility of this procedure by applying it to the study of doubly excited, for which conventional many-body methods struggle to describe. Within the CC framework, one often turns the excited-state equation-of-motion (EOM) extensions.\cite{EOMCCSD1,EOMCCSD2,EOMCCSD3,EOMCCSDTandTQ1,EOMCCSDTandTQ2,EOMCCSDT3,EOMCCSDT4,EOMCCSDT5,EOMCCSDTQ1,EOMCCSDTQ2} with higher-than-double excitations to describe these states, such as the EOM approach with single, double, and triple excitations (EOM-CCSDT)\cite{EOMCCSDTandTQ1,EOMCCSDTandTQ2,EOMCCSDT3,EOMCCSDT4,EOMCCSDT5} or up to quadruple excitations (EOM-CCSDTQ)\cite{EOMCCSDTandTQ1,EOMCCSDTandTQ2,EOMCCSDTQ1,EOMCCSDTQ2}. The full treatment of triple excitations with the EOM-CCSDT and EOM-CCSDT approaches have $\mathcal{N}^8$ and $\mathcal{N}^{10}$ computational steps, respectfully, where $\mathcal{N}$ is a measure of the system size. In contrast, the DUCC approaches in this paper have $\mathcal{N}^6$ steps when the ground-state CCSD approach is used as a source of amplitudes while avoiding complications associated with excited-state algorithms. The final reduced dimensionality of each problem, expressed by the downfolded Hamiltonian, represents system sizes amenable to current and near-term quantum computing architecture.

\section{\label{sec:Theory} Theory}

\subsection{Canonical CC Theory}
Determining the excitation energy of an excited state using a non-Aufbau reference determinant requires two separate calculations. The first uses a standard Aufbau reference corresponding to the ground state, denoted $\ket{\Phi_{0}}$. The second calculation uses the non-Aufbau reference determinant that correlates with the $\mu$-th excited state, denoted $\ket{\Phi_{\mu}}$. One can take the energy difference at this stage, which is referred to as the $\Delta$-SCF approximation. These states' descriptions are then improved by post-SCF calculations using suitable methods that treat these states separately. The exponential ansatz of coupled cluster theory is particularly well-suited for such calculations. 

The single-reference CC theory utilizes the exponential representation of the wave function $\ket{\Psi_{\mu}}$ 
    \begin{equation}
        \ket{\Psi_{\mu}} = e^{T_{\mu}}\ket{\Phi_{\mu}} \;,
        \label{CC-ansatz}
    \end{equation}
where $T_{\mu}$ is the cluster operator and $\ket{\Phi_{\mu}}$ is the single-determinantal reference function corresponding to either the ground state ($\mu = 0$) or an
excited state ($\mu > 0$). The cluster operator can be represented through its many-body components 
$T_{\mu,n}$
    \begin{equation}
        T_{\mu} = \sum_{n=1}^M T_{\mu,n} \;,
        \label{T-Operator}
    \end{equation}
where the individual component $T_{\mu,n}$ take the form
    \begin{equation}
        T_{\mu,n} = \frac{1}{(n!)^2} \sum_{\substack{i_1,\ldots,i_n \\ a_1,\ldots, a_n}} t^{i_1\ldots i_n}_{\mu,a_1\ldots a_n} E^{a_1\ldots a_n}_{i_1\ldots i_n} \;.
        \label{T-Cluster}
    \end{equation}
In the expressions, indices $i_1,i_2,\ldots$ ($a_1,a_2,\ldots$) refer to occupied (unoccupied) spin orbitals in the reference function $\ket{\Phi_{\mu}}$.
The excitation operators $E^{a_1\ldots a_n}_{i_1\ldots i_n} $ are defined through strings of standard creation ($a_p^{\dagger}$) and annihilation ($a_p$)
operators
    \begin{equation}
        E^{a_1\ldots a_n}_{i_1\ldots i_n}  = a_{a_1}^{\dagger}\ldots a_{a_n}^{\dagger} a_{i_n}\ldots a_{i_1} \;,
    \end{equation}
where creation and annihilation operators satisfy conventional anti-commutation rules.
When $M$ in the summation in Eq.~\ref{T-Operator} is equal to the number of correlated electrons ($N_e$), then the corresponding CC formalism is equivalent to the exact, full configuration interaction (FCI) solution, while truncations {($M<N_e$)} lead to the hierarchy of standard CC approximations such as CC with singles and doubles (CCSD, $M=2$), CC with singles, doubles, and triples (CCSDT, $M=3$), and so on. 

The amplitudes $t^{i_1\ldots i_n}_{\mu,a_1\ldots a_n}$ in Eq.~\ref{T-Cluster} are determined by solving a coupled system of energy-independent non-linear algebraic equations 
    \begin{equation}
         \matrixel{\Phi^{a_1\ldots a_n}_{\mu,i_1\ldots i_n}}{\overline{H}_{\mu}}{\Phi_{\mu}} \;,
    \end{equation}
corresponding to the amplitudes that are being solved, where $n = 1,\ldots,M$, 
    \begin{equation}
         \overline{H}_{\mu} = e^{-T_{\mu}}He^{T_{\mu}} = (He^{T_{\mu}})_C  \;,
         \label{SimH}
    \end{equation}
is the similarity-transformed Hamiltonian, $C$ designates the connected part of the operator expression, and $\ket{\Phi^{a_1\ldots a_n}_{\mu,i_1\ldots i_n}} = a_{a_1}^{\dagger}\ldots a_{a_n}^{\dagger} a_{i_n}\ldots a_{i_1}\ket{\Phi_{\mu}}$ are the $n$-tuply excited determinants relative to the reference $\ket{\Phi_{\mu}}$. Once the amplitudes are solved, the CC energy is determined using the equation
    \begin{equation}
         E_{\mu} =  \matrixel{\Phi_{\mu}}{\overline{H}_{\mu}}{\Phi_{\mu}} \;.
    \end{equation}
The excitation energy for the $\mu$-th excited state $\Delta E_{\mu}$ is the difference between the excited state energy $E_{\mu}$ ($\mu > 0$) and the ground state energy $E_{0}$.
    \begin{equation}
        \Delta E_{\mu} = E_{\mu} - E_{0} \;.
    \end{equation}
This procedure is referred to as the  $\Delta$CC method.


\subsection{DUCC Approach}
The DUCC formalism was developed for meaningful quantum chemistry calculations on limited quantum computing resources. The method is based on constructing effective (or downfolded) Hamiltonians in a small-dimensionality sub-space of the entire Hilbert space. While the DUCC approach is based on coupled cluster theory, one can not utilize the standard CC ansatz, given by Eq.~\ref{CC-ansatz}, as it leads to similarity-transformed Hamiltonians, Eq.~\ref{SimH}, that are non-Hermitian. Instead, an active space is used to introduce the DUCC ansatz that explicitly decouples excitations describing correlation effects inside (internal) and outside (external) of an active space, i.e.,
    \begin{equation}
        \ket{\Psi_{\mu, {\rm DUCC}}}=e^{\sigma_{\mu, {\rm ext}}} e^{\sigma_{\mu, {\rm int}}}\ket{\Phi_{\mu}} \;,
        \label{DUCC-ansatz}
    \end{equation}
where $\sigma_{\mu, {\rm int}}$ and $\sigma_{\mu, {\rm ext}}$ are anti-Hermitian cluster operators 
    \begin{eqnarray}
        \sigma_{\mu, {\rm int}}^{\dagger}&=&-\sigma_{\mu, {\rm int}} \;, \\
        \sigma_{\mu, {\rm ext}}^{\dagger}&=&-\sigma_{\mu, {\rm ext}} \;.
    \end{eqnarray} 
The algebraic form of the exact 
$\sigma_{\mu, {\rm int}}$ and $\sigma_{\mu, {\rm ext}}$ operators can effectively be approximated using UCC formalism, i.e., 
    \begin{eqnarray}
        \sigma_{\mu, {\rm int}} &\simeq& T_{\mu, {\rm int}}-T^{\dagger}_{\mu, {\rm int}} \;, \\
        \sigma_{\mu, {\rm ext}} &\simeq& T_{\mu, {\rm ext}}-T^{\dagger}_{\mu, {\rm ext}} \;.
    \end{eqnarray}
Once again, the $\mu$ notation is used because the DUCC approach can describe excited states with a proper reference determinant choice despite starting as a ground-state formalism. 

The DUCC formalism is described in greater detail in Refs. \onlinecite{DUCC1,DUCC_dynamics,bauman2022coupled}, but the foundation of the downfolding approach is that the energy $E_{\mu}$ can be obtained by diagonalizing an effective Hamiltonian $\bar{H}_{\rm ext}^{\rm eff(DUCC)}$ in the corresponding active space (defined by projection operator $P+Q_{\rm int}$, where $P$ and $Q_{\rm int}$ are projection operators onto the reference function and electron-promoted determinants in the active space, respectively) 
    \begin{equation}
        \bar{H}_{\mu}^{\rm eff(DUCC)} e^{\sigma_{\mu, {\rm int}}} \ket{\Phi_{\mu}} = E_{\mu} e^{\sigma_{\mu, {\rm int}}}\ket{\Phi_{\mu}} \;,
        \label{DUCC-eig}
    \end{equation}
where
    \begin{equation}
        \bar{H}_{\mu}^{\rm eff(DUCC)} = (P+Q_{\mu, {\rm int}}) \bar{H}_{\mu, {\rm ext}}^{\rm DUCC} (P+Q_{\mu, {\rm int}})
    \end{equation}
and
    \begin{equation}
        \bar{H}_{\mu, {\rm ext}}^{\rm DUCC} =e^{-\sigma_{\mu, {\rm ext}}}H e^{\sigma_{\mu, {\rm ext}}} \;.
    \end{equation}
In practice, one constructs approximate many-body forms of $\bar{H}_{\mu, {\rm ext}}^{\rm eff(DUCC)}$. 

Three considerations go into the approximation of $\bar{H}_{\mu, {\rm ext}}^{\rm eff(DUCC)}$: 
(1) the rank of many-body effects included in $\bar{H}_{\rm ext}^{\rm eff(DUCC)}$, (2) the 
approximate representation of $\sigma_{\mu, {\rm ext}}$ ($T_{\mu, {\rm ext}}$), and (3) the 
length of the commutator expansion for $e^{-\sigma_{\rm ext}}H e^{\sigma_{\rm ext}}$. In this 
paper, I limit the effective Hamiltonian to one- and two-body elements. I also employ 
traditional CC theory to source the approximate $T_{\rm ext}$ operators that parameterize 
$\sigma_{\rm ext}$ ($\sigma_{\mu, {\rm ext}} \simeq T_{\mu, {\rm ext}}-T^{\dagger}_{\mu, {\rm ext}}$). 
Finally, in a previous study, we showed that approximating the expansion of  $e^{-
\sigma_{\rm ext}}H e^{\sigma_{\rm ext}}$ so as to be consistent through second-order 
perturbation theory labeled DUCC(2), given by 
    \begin{equation}
        \bar{H}_{\mu, {\rm ext}}^{\rm DUCC(2)} \simeq H+[H_N,\sigma_{\mu, {\rm ext}}]+\frac{1}{2!}[[F_N,\sigma_{\mu, {\rm ext}}],\sigma_{\mu, {\rm ext}}] \;,
        \label{2nd-order}
    \end{equation}
performs well in describing the system with the third-order consistent expression labeled DUCC(3) and given by
    \begin{eqnarray}
        \bar{H}_{\mu, {\rm ext}}^{\rm DUCC(3)} \simeq &&H+[H_N,\sigma_{\mu, {\rm ext}}]+\frac{1}{2!}[[H_N,\sigma_{\mu, {\rm ext}}],\sigma_{\mu, {\rm ext}}] + \nonumber \\
        &&\frac{1}{3!}[[[F_N,\sigma_{\mu, {\rm ext}}],\sigma_{\mu, {\rm ext}}],\sigma_{\mu, {\rm ext}}] \;,
        \label{3rd-order}
    \end{eqnarray}
further improving upon the results. In Eqs.~\ref{2nd-order} and \ref{3rd-order}, $F_N$ and $H_N$ are the normal-product Fock and Hamiltonian operators. 

\section{\label{sec:Results} Results and Discussion}
I investigated the doubly excited states of four molecules: H$_{2}$, Methylene, Formaldehyde, and Nitroxyl. For H$_{2}$, the cc-pVTZ basis set was employed, while aug-cc-pVDZ was used for Methylene, Formaldehyde, and Nitroxyl. The $T_{\mu, {\rm ext}}$ amplitudes defining $\sigma_{\mu, {\rm ext}}$ are sourced from traditional CCSD ($\sigma_{\mu, {\rm ext}} \simeq T^{\rm (CCSD)}_{\mu, {\rm ext}}-T^{\rm (CCSD)\dagger}_{\mu, {\rm ext}}$). Any method that utilizes a non-Aufbau reference determinant to describe an excited state is given the prefix "ES-". Lastly, excitation energies computed as the difference between "ES-" excited state energies and the canonical ground state energies using their corresponding reference determinants are labeled with the prefix "$\Delta$-". Core orbitals were frozen in all post-SCF calculations of Formaldehyde and Nitroxyl. In the case of the DUCC methods, core orbitals were frozen after forming the effective Hamiltonian from all-electron calculations.

\subsection{\label{sec:H2}H$_{2}$}

\begin{figure}
        \includegraphics[width=0.45\textwidth]{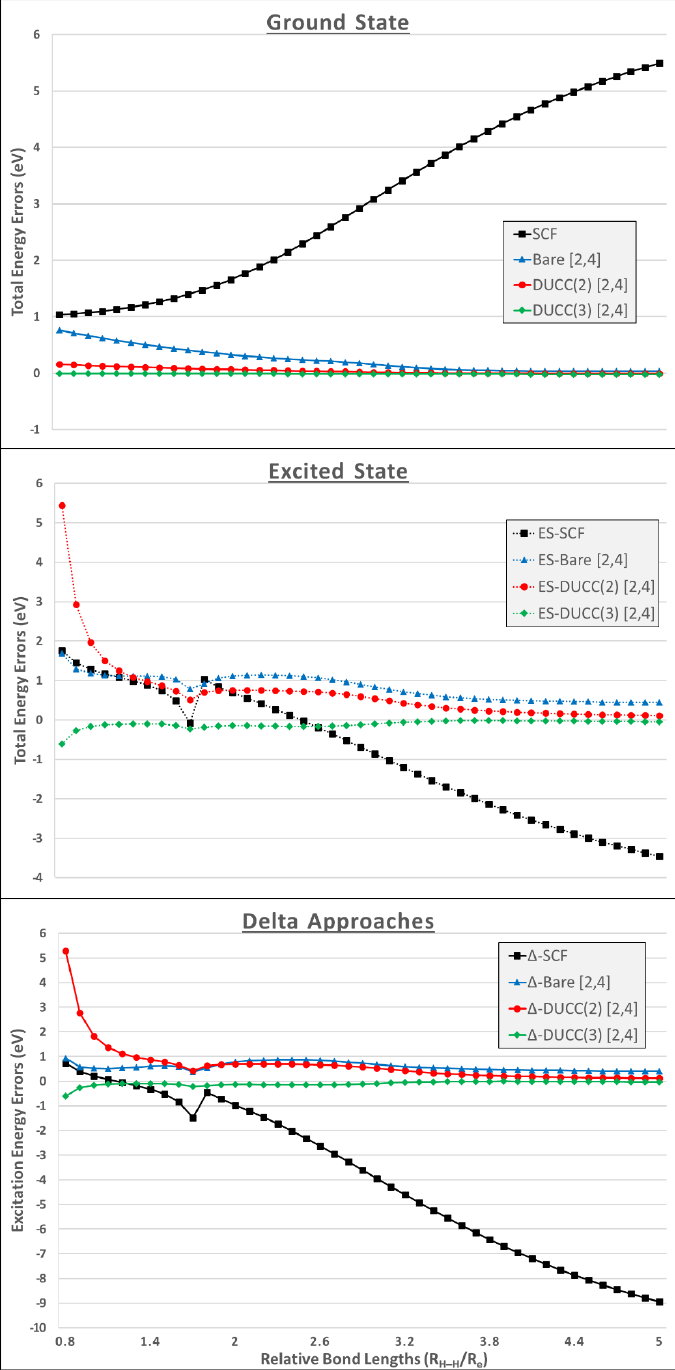}
        \caption{\label{fig:H2} Errors relative to FCI for the ground state (Top Panel), [(1$\sigma_{g}$)$^{2} \rightarrow $ (1$\sigma_{u}$)$^{2}$] excited state (Middle Panel), and the corresponding excitation energy (Bottom Panel) along the $H$--$H$ bond-breaking coordinate. The equilibrium bond length is taken to be 0.7414 \AA. }
    \end{figure}

We start our discussion by investigating the (1$\sigma_{g}$)$^{2} \rightarrow $ (1$\sigma_{u}$)$^{2}$ excitation along the $H$--$H$ bond-breaking coordinate. In this case, CCSD is equivalent to FCI, so using CCSD as a source of amplitudes to define $\sigma_{\mu, {\rm ext}}$ removes any concern about the role of higher-order excitations. An active space constructed from the four lowest-energy orbitals was used for the downfolding procedure, where the full space consists of 28 orbitals. The results are shown in Fig.~\ref{fig:H2}.

The $\Delta$-SCF is praised for providing excitation energies similar to FCI near equilibrium, but that is because of fortuitous balanced errors in the total energies for the ground and excited state, which is not the case for other geometries along the potential energy surface. Diagonalizing the bare Hamiltonian in the active space provides a much-needed improvement, especially at stretched bond lengths, but it is not balanced for both states. When correlation is downfolded with the DUCC(2) approach, total energies for the ground state agree with FCI to within 0.16 eV along the whole potential energy surface and improve upon the bare Hamiltonian for the excited state with the exception of significantly compressed bond lengths. When the DUCC(3) method is utilized, all energies are improved compared to the DUCC(2) approach, and one gets both total and excitation energies that agree excellently with FCI results by diagonalizing the effective Hamiltonian in an active space of only four orbitals.

\subsection{\label{sec:Methylene}Methylene}
The first excited state (1 $^{1}A_{1}$) and third excited state (2 $^{1}A_{1}$) of Methylene have considerable mixing of configurations $(1a_{1})^{2}(2a_{1})^{2}(1b_{1})^{2}(3a_{1})^{2}$ and $(1a_{1})^{2}(2a_{1})^{2}(1b_{1})^{2}(1b_{2})^{2}$. The (1 $^{1}A_{1}$) state can be computed using ground state methods as it is the lowest-energy singlet state. The (2 $^{1}A_{1}$) state can then be calculated as an excited state formed from a double excitation out of the (1 $^{1}A_{1}$) state. The multiconfigurational character of the (1 $^{1}A_{1}$), while non-negligible, is minor enough that ground-state methods such as CCSD can provide an accurate description of the state. However, the mixing of configurations is much more prominent in the (2 $^{1}A_{1}$). When calculated as a doubly excited state from (1 $^{1}A_{1}$), one needs to include higher-than-double excitations to capture this state accurately. 

\begin{figure}
        \includegraphics[width=0.44\textwidth]{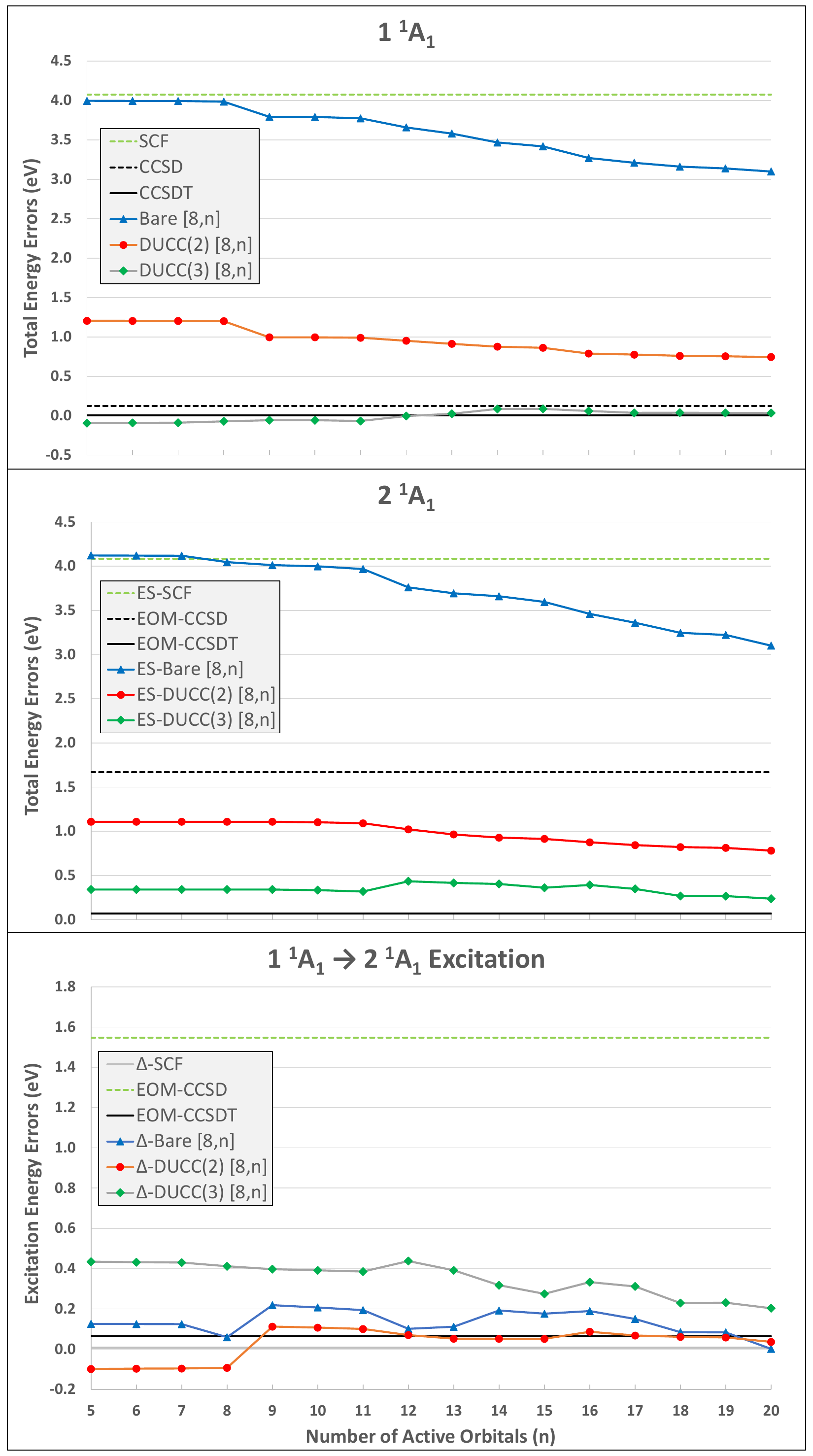}
        \caption{\label{fig:Methylene} Errors relative to CCSDTQ and EOMCCSDTQ for the (1 $^{1}A_{1}$) state (Top Panel), (2 $^{1}A_{1}$) excited state (Middle Panel), and the corresponding excitation energy (Bottom Panel) of the methylene molecule. The $C-H$ bond length was 1.107$\text{\AA}$, and the $H-C-H$ angle was 102.4 degrees. }
    \end{figure}

As shown in Fig. \ref{fig:Methylene}, while CCSD and CCSDT provide accurate results of the (1 $^{1}A_{1}$) state, EOM-CCSD struggles to describe the (2 $^{1}A_{1}$) state, with an error of 1.67 eV. This error vanishes with EOM-CCSDT, which is nearly exact for this system. For the downfolding procedure, several active spaces were explored consisting of 5-20 of the lowest-energy molecular orbital of the Aufbau and non-Aufbau reference determinants (denoted as $n$), whereas the full orbital space consists of 41 orbitals. The DUCC(2) approximation substantially improves the total energies over the bare Hamiltonian in the same size active space. Both states' DUCC(2) energies are well-balanced, leading to excitation energies that agree well with the benchmark EOM-CCSDTQ result. However, there is still room for improvement in the total energies with the DUCC(2) approach. Higher-order terms in the commutator expansion play a crucial role in providing accurate total energies for the two states, as shown with the DUCC(3) approach. For the (1 $^{1}A_{1}$) state, the DUCC(3) approach provides total energies between CCSD and CCSDT, while for the (2 $^{1}A_{1}$) state it reduces the errors to less than 0.44 eV for any active space. The state-specific nature of the downfolding procedure means that the errors are not always balanced. between given states, which is why the DUCC(3) approach has larger errors in the excitation energy than the DUCC(2) approximation. Still, there is a substantial improvement with the DUCC(3) procedures over EOM-CCSD in describing this excitation with a steady improvement to the EOM-CCSDTQ benchmark as the active-space size increases.

\subsection{\label{sec:Formaldehyde}Formaldehyde}

Formaldehyde has a doubly excited state [(2$^{1}B_{1}$)$^{2}$ $ \rightarrow $ (2$^{1}B_{2}$)$^{2}$] that has been troublesome for many low-level many-body methods. The EOM-CCSD method overestimates the excitation energy by 4 eV. The EOM-CCSDT method greatly improves upon this excitation energy, but one needs to include quadruple excitations through EOM-CCSDTQ to get nearly exact results. The ground state is well-described in the case of CCSD and CCSDT. It is the excited state description that leads to errors in excitation energy. When the non-Aufbau reference determinant is used, the ES-CCSD result provides excited-state total energy comparable to canonical EOM-CCSDT, while the ES-CCSDT method provides canonical EOM-CCSDTQ quality results. The agreement of the ES-CCSD and ES-CCSDT methods reinforces the notion of using the [(2$^{1}B_{1}$)$^{2}$ $ \rightarrow $ (2$^{1}B_{2}$)$^{2}$] determinant as the reference for the downfolding procedure.

\begin{figure}
        \includegraphics[width=0.44\textwidth]{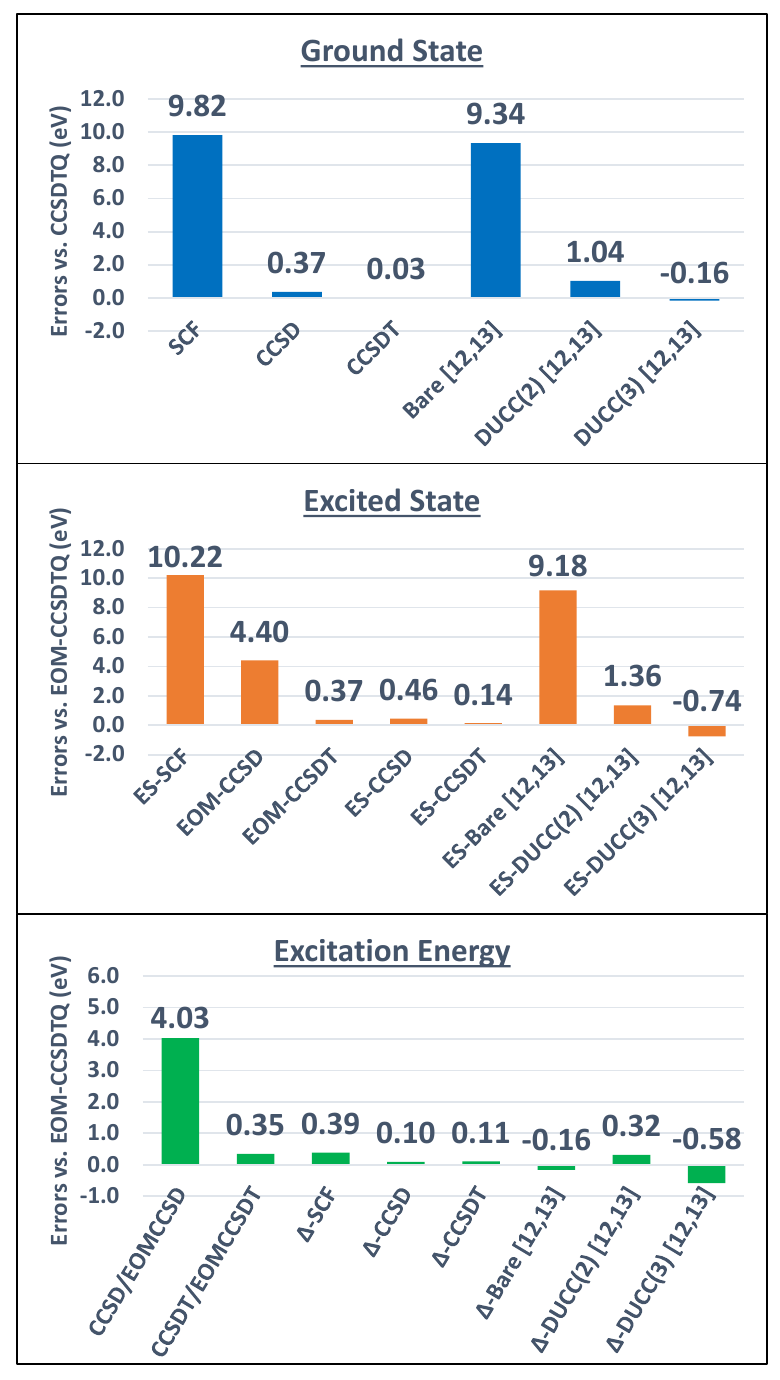}
        \caption{\label{fig:Formaldehyde} Errors relative to CCSDTQ and EOMCCSDTQ for the ground state (Top Panel), [(2$^{1}B_{1}$)$^{2}$ $ \rightarrow $ (2$^{1}B_{2}$)$^{2}$] excited state (Middle Panel), and the corresponding excitation energy (Bottom Panel) of the formaldehyde molecule. The equilibrium geometry was taken from Ref. \onlinecite{loos2019reference}.}
    \end{figure}

For the downfolding procedure, an active space consisting of the 15 (13 after freezing the core orbitals) lowest-energy molecular orbitals for both the Aufbau and non-Aufbau reference determinants was utilized. The full calculation involves 64 orbitals. Through the maximum overlap method, I confirmed that all of the orbitals in the Aufbau determinant have corresponding analogous orbitals in the non-Aufbau reference determinant and that different orbitals are not introduced into the active space during the SCF procedure. Illustrated in Fig.~\ref{fig:Formaldehyde}, the bare Hamiltonian provides excellent excitation energy but at the cost of over 9 eV errors in both the ground and excited state. With only 13 orbitals, the DUCC(2) approach greatly improved both total energies and provided EOM-CCSDT quality excitation energy. The DUCC(3) approach further improves the total energies, while a slight imbalance in errors leads to a small increase in the error of the excitation energy. Both the DUCC(2) and DUCC(3) approaches provide an impressive improvement compared to EOM-CCSD in describing the excited state, with the DUCC(3) method providing total energies for both states comparable to high-level CC approaches. 

\subsection{\label{sec:Nitroxyl}Nitroxyl}

Nitroxyl is a weak acid that has a low-lying [(7$^{1}a'$)$^{2}$ $ \rightarrow $ (2$^{1}a''$)$^{2}$] excited state. In a similar fashion to Formaldehyde, the EOM-CCSD method overestimates the excitation energy by 4 eV. The EOM-CCSDT method reduces the excitation energy error to ~0.3 eV, and once again, quadruple excitations through EOM-CCSDTQ are needed to get nearly exact results. For the EOM-CCSD and EOM-CCSDT approaches, the errors in the excited state are approximately 10 times greater than the ground state, which contributes to the excitation energy errors. However, when the non-Aufbau reference determinant is used, once again, the ES-CCSD result is comparable to canonical EOM-CCSDT, and the ES-CCSDT method provides canonical EOM-CCSDTQ-level results. So the [(7$^{1}a'$)$^{2}$ $ \rightarrow $ (2$^{1}a''$)$^{2}$] reference determinant was used in the ground-state DUCC pipeline.

\begin{figure}
        \includegraphics[width=0.44\textwidth]{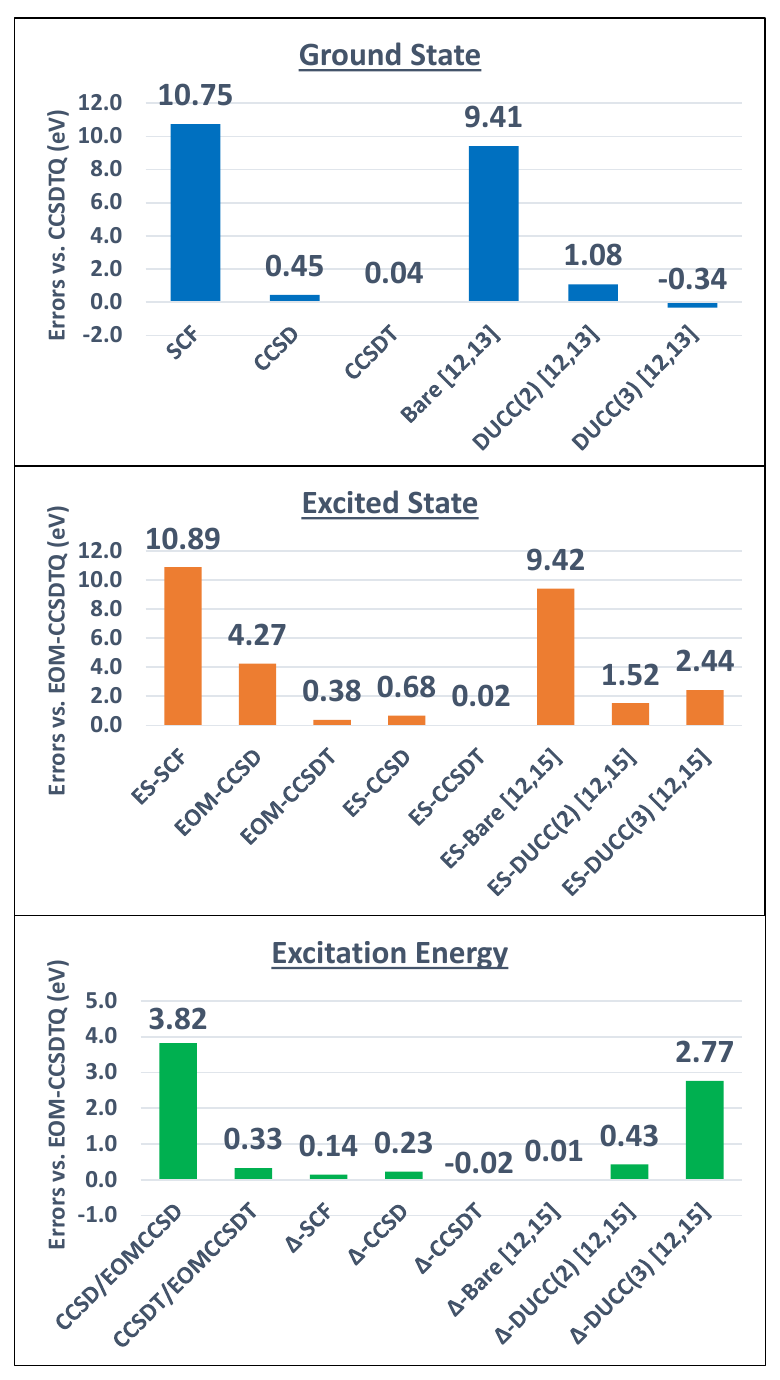}
        \caption{\label{fig:Nitroxyl-1} Errors relative to CCSDTQ and EOMCCSDTQ for the ground state (Top Panel), [(7$^{1}a'$)$^{2}$ $ \rightarrow $ (2$^{1}a''$)$^{2}$] excited state (Middle Panel), and the corresponding excitation energy (Bottom Panel) of the nitroxyl molecule. The equilibrium geometry was taken from Ref. \onlinecite{loos2019reference}.}
    \end{figure}

An active space consisting of the 17 (15 after freezing the core orbitals) lowest-energy molecular orbitals for both the Aufbau and non-Aufbau reference determinants was used, whereas the full calculation contains 55 molecular orbitals. Through the maximum overlap method, I confirmed that all of the orbitals in the Aufbau determinant have corresponding analogous orbitals in the non-Aufbau reference determinant. As shown in Fig.~\ref{fig:Nitroxyl-1}, diagonalizing the truncated bare Hamiltonian gives nearly exact excitation energy, but ground and excited states have total energy errors over 9 eV. Both the DUCC(2) and DUCC(3) approaches greatly improve the total energies in the active space compared to the bare Hamiltonian. However, his improvement is not always balanced and can lead to excitation energies with undesired errors. Since either approach significantly improves the EOM-CCSD total energy for the excited state, the DUCC(2) and DUCC(3) approximations still provide better excitation energy than EOM-CCSD despite the imbalance total energy errors. 

\begin{figure}
        \includegraphics[width=0.44\textwidth]{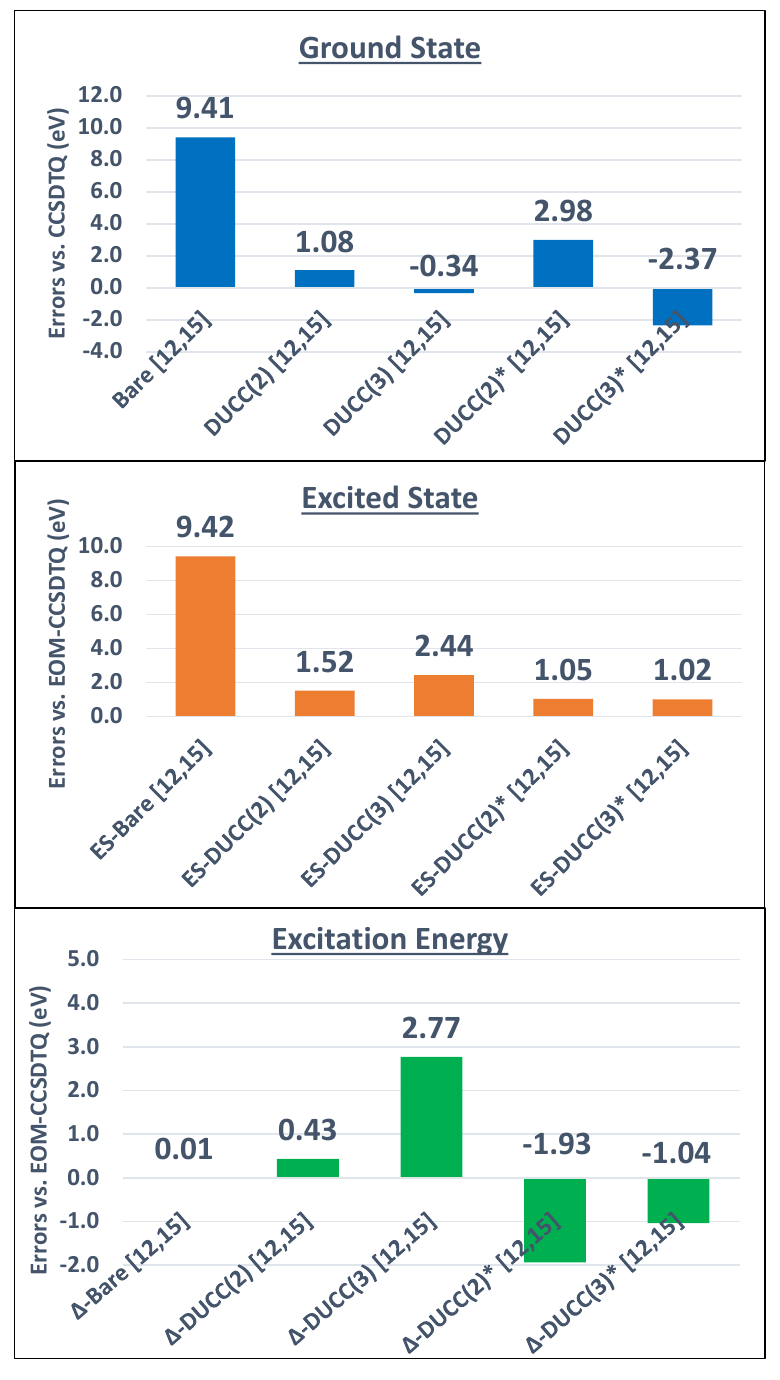}
        \caption{\label{fig:Nitroxyl-2} Comparing the source of amplitudes defining the approximate $\sigma_{\mu, {\rm ext}}$ operator. When CCSDT is used as a source of amplitudes, the method is denoted with an asterisk *. Errors are relative to CCSDTQ and EOMCCSDTQ for the ground state (Top Panel), [(7$^{1}a'$)$^{2}$ $ \rightarrow $ (2$^{1}a''$)$^{2}$] excited state (Middle Panel), and the corresponding excitation energy (Bottom Panel) of the nitroxyl molecule. The equilibrium geometry was taken from Ref. \onlinecite{loos2019reference}.}
    \end{figure}

Compared to Formaldehyde, Nitroxyl exhibits a greater difference between CCSD and CCSDT errors for the ground state and ES-CCSD and ES-CCSDT errors for the excited state. Between ES-CCSD and ES-CCSDT, the energy difference is 0.66 eV. Upon investigation, there is a $T_2$ amplitude in the ES-CCSD calculation with a value of 0.61, which is less than 0.1 in the ES-CCSDT calculation. Given the largest cluster amplitude fluctuation, we performed the downfolding procedure with $T_1$ and $T_2$ amplitudes from the corresponding CCSDT and ES-CCSDT calculations as terms with explicit $T_3$ contributions have not been coded yet. Results in Fig.~\ref{fig:Nitroxyl-2} show that higher-order excitations play a non-negligible role in defining the effective Hamiltonian for Nitroxyl, changing total energies by up to 2 eV. In the case of the excited state, we notice a positive improvement in total energies, but the same cannot be said for the ground state. However, for the time being, this story is left incomplete because if triple excitations play such a significant role, it stands to reason those explicit $T_3$ (and higher-order) contributions, longer commutator expansions, and higher-order many-body elements of the Hamiltonian can all endow further changes to the description of the ground and excited states. 

\subsection{Overall}
Through the four examples, several trends can be observed. First, downfolding procedures with non-Aufbau references can accurately describe excited states in compact active spaces with substantial improvements over the active-space bare Hamiltonian and even EOM-CCSD for the representative difficult cases studied. Second is that higher-order terms in the commutator expansion play a significant role in improving accuracies of total energies, as seen with the difference between the DUCC(2) and DUCC(3) approaches. Third, since the downfolding procedure is state state-specific, the description between two states is not always balanced. It is important to reiterate that accurate excitation energies in and method may be fortuitous at the expense of total energies. The downfolding procedures in this paper aim to provide accurate total energies and consequently excitation energies. To further improve upon the total energies and provide a balanced description, the role of longer commutator expansions, higher-order excitations, and higher body terms in the downfolded Hamiltonian needs to be investigated.

As seen with methylene, the description of the two states systematically improves as the active space size increases. The active spaces of the remaining molecules represent system sizes well-suited for current and near-term quantum computing architectures. Larger active spaces can be explored as quantum computers grow in their quantum capacity. This means that the approximations in the downfolding procedure are more befitting, and the description of states becomes more accurate and balanced. One can imagine reliably downfolding 100s or 1000s of orbitals to active spaces that are magnitudes of order smaller, greatly expanding the envelope of problems one can tackle with quantum computers.

\section{Conclusion}
In this paper, I explored how ground-state downfolding coupled cluster techniques are efficient tools that can target excited 
states using non-Aufbau reference determinants. The double unitary coupled cluster ansatz employed produces state-specific hermitian effective Hamiltonians of reduced dimensionality corresponding to an active space used to partition the excitations in the ansatz. The final reduced dimensionality of each problem, expressed by the downfolded Hamiltonian, represents system sizes amenable to current and near-term quantum computing architecture.

We focused on doubly excited states of H$_{2}$, Methylene, Formaldehyde, and Nitroxyl, and demonstrated that downfolding techniques result in state-specific effective Hamiltonians that, when diagonalized in their respective active spaces, provide ground- and excited-state total energies (and therefore excitation energies) comparable to high-level CC methods. The downfolding procedures consistently improve the accuracy of total energies compared to the bare Hamiltonian within the same active space size. The downfolding procedures also provide excited state energies that outperform EOM-CCSD, leading to better total and excitation energies. This technique of combining non-Aufbau reference determinants with downfolding procedures allows for accurate and reliable investigation of excited states with a significant reduction in dimensionality. Together with excited-state quantum computing techniques and algorithms, this technique greatly broadens the envelope of excited-state quantum chemistry problems for quantum computers.

\section{Acknowledgement}
This work was supported the Quantum Science Center (QSC), a National Quantum Information Science Research Center of the U.S. Department of Energy (under
FWP 76213) and the Laboratory Directed Research and Development (LDRD) Program at PNNL. This work used resources from the Pacific Northwest National Laboratory
(PNNL). PNNL is operated by Battelle for the U.S. Department of Energy under Contract DE-AC05-76RL01830.

\nocite{*}
\bibliography{aipsamp}

\end{document}